\documentstyle[12pt]{article}
\def \ms {{\overline{\mbox{MS}}}}

\newcommand{\titul}[1] {\begin{center}{\large\bf #1 } \end{center}\vskip 1.cm}
\newcommand{\autor}[1] {\begin {center} {\large \lineskip .5em #1 }
                        \end   {center} }
\newcommand{\lugar}[1] {\begin{center} {\it #1} \end{center}}
\newcommand{\abstr}[1] {{\begin{center} \vskip .5cm {\bf Abstract
                        \vspace{0pt}} \end{center}}\begin{quote} #1
                        \end{quote}}
\begin{document}
%\begin{titlepage}
.
%\begin{flushright} {\bf US-FT/13-98 \\ July 4th, 1998} \end{flushright}

\vskip -1.cm
\titul{ $Q^2$ evolution of parton distributions\\ at low
$x$. Soft initial conditions
}

\autor{A.V. Kotikov
%\footnote{E-mail:KOTIKOV@SUNSE.JINR.DUBNA.SU}
}
\lugar{Particle Physics Laboratory\\
Joint Institute for Nuclear Research\\
141980 Dubna, Russia}
\autor{G. Parente
%\footnote{E-mail:GONZALO@GAES.USC.ES}
}
\lugar{Departamento de F\'\i sica de Part\'\i culas\\
Universidade de Santiago de Compostela\\
15706 Santiago de Compostela, Spain}
\abstr{
We investigate  the  $Q^2$ evolution of parton distributions
at small $x$ values,
%present an analytical parametrization of the QCD
%description of the  behaviour
%of parton distribution functions in the leading twist approximation
%of the Wilson operator product expansion,
recently obtained in the case
of soft initial conditions.
The results are in excellent agreement with deep inelastic scattering
experimental data from HERA.

%PACS number(s): 13.60.Hb, 12.38.Bx, 13.15.Dk

}
%
%\end{titlepage}
%\newpage

%\pagestyle{plain}

%\section{Introduction} \indent

The measurements of the deep-inelastic scattering
%(DIS)
structure function
%(SF)
$F_2$ in HERA
%\cite{H1,ZEUS,ZEUSB}
\cite{H1}
have permitted the access to
a very interesting kinematical range for testing the theoretical
ideas on the behavior of quarks and gluons carrying a very low fraction
of momentum of the proton, the so-called small $x$ region.
In this limit one expects that
non-perturbative effects may give essential contributions. However, the
resonable agreement between HERA data and the NLO approximation of
perturbative
QCD that has been observed for $Q^2 > 1 $GeV$^2$ (see the recent review
in \cite{CoDeRo}) indicates that
%and, thus,
perturbative QCD could describe the
evolution of structure functions up to very low $Q^2$ values,
traditionally explained by soft processes.
It is of fundamental importance to find out the kinematical region where
the well-established perturbative QCD formalism
can be safely applied at small $x$.

The standard program to study the small $x$ behavior of
quarks and gluons
is carried out by comparison of data
with the numerical solution of the
%Dokshitzer-Gribov-Lipatov-Altarelli-Parisi (DGLAP)
DGLAP
equations 
%\cite{DGLAP1, DGLAP}
\footnote{ At small $x$ there is another approach
based on the
%Balitsky-Fadin-Kuraev-Lipatov (BFKL) equation \cite{BFKL},
BFKL equation, whose
application is out of the scope of this work.} by
fitting the parameters of the
$x$ profile of partons at some initial $Q_0^2$ and
the QCD energy scale $\Lambda$ (see, for example, \cite{MRS}).
%\cite{fits}-\cite{GRV}.
However, if one is interested in analyzing exclusively the
small $x$ region, there is the alternative of doing a simpler analysis
by using some of the existing analytical solutions of DGLAP 
in the small $x$ limit (see \cite{CoDeRo} for review).
%\cite{BF1}-\cite{Munich2}.
This was done so in Ref. \cite{BF1}-\cite{Q2evo}
where it was pointed out that the HERA small $x$ data can be
interpreted in 
terms of the so called doubled asymptotic scaling phenomenon
related to the asymptotic 
behaviour of the DGLAP evolution 
discovered  in \cite{Rujula}
many years ago. Here we will illustrate results obtained recently in
\cite{Q2evo}.\\

{\bf 1.} Thus, our purpose
%of this article
is to demonstrate the small $x$ asymptotic
form of parton distributions
%(PD)
in the framework of the DGLAP equation starting at some $Q^2_0$ with
the flat function:
 \begin{eqnarray}
f_a (Q^2_0) ~=~
A_a ~~~~(a=q,g), \label{1}
 \end{eqnarray}
where $f_a$ are the parton distributions multiplied by $x$
and $A_a$ are unknown parameters that have to be determined from data.
Through this work at small $x$ we neglect
the non-singlet quark component.

In \cite{Q2evo} an effective method to reproduce the $x$-dependence of
parton distributions has been developed. It is based on a separation
of the singular and regular parts of the  exact solution for the
moments of
parton distributions. The method allows in simplest way to reproduce the
leading order (LO) results \cite{BF1} and to construct the $x$-dependence
of parton distributions at next-to-leading order (NLO):
% It has the form:
 \begin{eqnarray}
 f_a(x,Q^2) &=& f_a^+(x,Q^2) + f_a^-(x,Q^2)~~~ ~\mbox{ and }~ \nonumber \\
f_a^-(x,Q^2)&=& A_a^-(Q^2,Q^2_0) exp(- d_{-}(1)s-D_{-}(1)p) ~+~O(x) 
%\label{9.10} \\
\nonumber \\
f_g^+(x,Q^2) &=& A_g^+(Q^2,Q^2_0)
%&\cdot & 
I_0(\sigma) exp(- \overline d_{+}(1)s-\overline D_{+}(1)p)
~+~O(\rho) 
%~~\mbox { and } 
%\nonumber \\
\label{9.11} \\
f_q^+(x,Q^2)&=& 
%A_q^-  exp(- d_{-}(1)s-D^q_{-}(1)p) 
%\label{9.2} \\&+& 
A_q^+(Q^2,Q^2_0)
\biggl[ (1 - \bar{d}_{\pm}^q(1) \alpha(Q^2)) \rho I_1(\sigma)
      + 20 \alpha(Q^2) I_0(\sigma) \biggr]
\nonumber \\
& & 
%&\cdot &
\cdot 
exp(- \overline d_{+}(1)s-\overline D_{+}(1)p)
+O(\rho)
\nonumber \\
%\label{9.12} \\
%
F_2(x,Q^2)&=& e \cdot \biggl(f_q(x,Q^2) + \frac{2}{3}f\alpha(Q^2) f_g(x,Q^2)
 \biggr) \nonumber 
%\label{9}
\end{eqnarray}
where $I_{\nu}(\sigma)$ is modified Bessel function, $s$ and $p$
are given by 
$s=ln(\alpha (Q^2_0)/\alpha (Q^2)),
~p=\alpha (Q^2_0) - \alpha (Q^2)$, and 
% \begin{eqnarray}
$$\sigma = 2\sqrt{(\hat d_{+}s+\hat D_{+}p)lnx} ~~, ~~~~
\rho = \sqrt{\frac{(\hat d_{+}s+\hat D_{+}p)}{lnx}}=
\frac{\sigma }{2ln(1/x)}
~~$$ are NLO generalizations of the corresponding Ball-Forte variables and
\begin{eqnarray}
%\label{a1}\\
A_g^+(Q^2,Q^2_0) &=& \Bigr[1-\frac{80}{81}f\alpha(Q^2) \Bigr]A_g 
+ \frac{4}{9}\Bigl[1+3(1+\frac{1}{81}f)\alpha(Q^2_0) - 
\frac{80}{81}f  \alpha(Q^2)
\Bigr] A_q ~~, 
\nonumber \\
A_g^-(Q^2,Q^2_0) &=& A_g - A_g^+(Q^2,Q^2_0)  \nonumber \\
A_q^+ &=& \frac{f}{9}\biggl(A_g + \frac{4}{9} A_q \biggl)~~, ~~~
A_q^- = A_q - 20 \alpha(Q^2_0) A_q^+
\label{a2} \end{eqnarray}
%are magnitudes of $\pm$ components.
Wherever in this work we use the notation $\alpha = \alpha_s/(4\pi)$.
The nonzero components of the singular and regular parts of the terms $d_{\pm}$
and $D_{\pm}$ have the form:

 \begin{eqnarray}
\hat d_{+} &=& -\frac{12}{\beta_0} ~~, ~~~
\overline d_{+}(1) =1+ \frac{4}{3\beta_0}f  ~~, ~~~
d_{-}(1) =\frac{16}{27\beta_0}f ~~, \nonumber \\
\hat d_{++} &=& \frac{412}{27\beta_0}f ~~, ~~~
\hat d^q_{+-} = -20 ~~, ~~~
\overline d^g_{+-}(1) = \frac{80}{81}f~~, \nonumber \\
\overline d_{++}(1) &=& \frac{8}{\beta_0}
\biggl( 36 \zeta_3 + 33 \zeta_2 - \frac{1643}{12} +\frac{2}{9}f 
\Bigr[ \frac{68}{9} -4 \zeta_2 - \frac{13}{243}f \Big] \biggr)~~, \nonumber \\
\overline d^q_{+-}(1) &=& 
\frac{134}{3} -12 \zeta_2 - \frac{13}{81}f ~~, ~~~ 
\overline d^g_{-+}(1) = -3 \Bigl( 1+ \frac{f}{81} \Bigr),~~~~ ~~~
\nonumber \\
d_{--}(1) &=& \frac{16}{9\beta_0}
\biggl( 2 \zeta_3 - 3 \zeta_2 + \frac{13}{4} + f 
\Bigr[  4 \zeta_2 - \frac{23}{18} + \frac{13}{243}f \Big] \biggr)~~,
%\nonumber \\
%
%d^q_{-+}(1) &=& 0~~, ~~~
%
%.
 \label{9.3}
 \end{eqnarray}
where $\beta_0$ is the first coefficient of QCD $\beta$-function,
$\zeta_n$ are Euler $\zeta$-functions and $f$ is the number of
active quarks.

Looking carefully Eqs. (\ref{9.11}), we arrive to the following 
conclusions:
\begin{itemize}
\item Our NLO results coincide with the corresponding of Ball
and Forte in Ref. \cite{BF2} if one neglects the ``$-$'' component,
expands our NLO singular terms
$(\rho)^k I_{k+1}(\sigma)$ in the vicinity of the point
$\sigma = \sigma_{LO} $ and ignores the NLO regular terms (i.e. put  
$exp{(-\overline D_{+}(1)p)}=exp{(-D_{-}(1)p)}=1$ and cancel the terms 
proportional to $\alpha(Q^2_0)$ into the normalization factors $A_g^\pm$
and $A_q^-$). We think, however, that this expansion is not so correct
because it generates NLO corrections of the order of the LO terms.

%\item The negative sign of the NLO correction in  $\sigma$
%(see Eq. (\ref{a1}))  
%makes excellent the agreement of our result with the parametrization of $F_2$
%obtained by De Roeck and De Wolf \cite{DRDW}.
%Their result is very similar to our LO form
%of $f_q^+$ in Eq.(\ref{8.0}) if one replaces  $s_{LO} \to
%s_{LO}^{\delta}$ in the definition (\ref{2.5}) of $s_{LO}$. The
%value $\delta = 0.708$ has been obtained in the fit to H1 and ZEUS
%data. Due to $\delta <1$, it shows less $Q^2$-dependence
%than it is predicted by perturbative QCD at LO. This slower
%$Q^2$-dependence may be explained naturally by the negative NLO corrections
%to $\sigma$ obtained here.

\item
The behaviour of eqs. (\ref{9.11}) can mimic a power law shape
over a limited region of $x, Q^2$:
 \begin{eqnarray}
f_a(x,Q^2) \sim x^{-\lambda^{eff}_a(x,Q^2)}
 ~\mbox{ and }~
F_2(x,Q^2) \sim x^{-\lambda^{eff}_{F2}(x,Q^2)}
\nonumber    \end{eqnarray}
The quark and gluon effective slopes
 $\lambda^{eff}_a $
%= -\frac{d}{d \ln z} \ln f_a(z,Q^2)$
are reduced by the NLO terms that leads to the decreasing
of the gluon distribution at small $x$. For the quark case
it is not the case, because the normalization factor $A_q^+$ of the ``$+$'' 
component produces an additional contribution undampening as $\sim (lnx)^{-1}$.

\item
The gluon effective slope $\lambda^{eff}_g$ is larger than the quark slope
$\lambda^{eff}_q$, which is in excellent agreement with a recent MRS and 
GRV analyses \cite{MRS}. \\
Indeed,
%because $d/dlnx = d/dlnz$,
the effective slopes
have the
%form,
asymptotical values (at large $Q^2$):
 \begin{eqnarray}
\lambda^{eff,as}_g(x,Q^2) &=& 
\rho \frac{I_1(\sigma)}{I_0(\sigma)} \approx \rho - 
\frac{1}{4\ln{(1/x)}} 
\nonumber \\
\lambda^{eff,as}_q(x,Q^2) &=& 
\rho \cdot \frac{ I_2(\sigma) (1- \overline d^q_{+-}(1) \alpha(Q^2))
 + 20 \alpha(Q^2) I_1(\sigma)/\rho}{ I_1(\sigma) 
(1- \overline d^q_{+-}(1) \alpha(Q^2))
 + 20 \alpha(Q^2) I_0(\sigma)/\rho}
 \nonumber \\
&\approx & \biggl( \rho - \frac{3}{4\ln{(1/x)}} \biggr) \biggl(1- 
\frac{10\alpha(Q^2)}{(\hat d_+ s + \hat D_+ p)} \biggr)
\label{11.1} \\
\lambda^{eff,as}_{F2}(x,Q^2) 
&=& \lambda^{eff,as}_q(x,Q^2) 
\frac{ 1 + 6 \alpha(Q^2)/\lambda^{eff,as}_q(z,Q^2)}{ 1 + 
6 \alpha(Q^2)/\lambda^{eff,as}_g(z,Q^2)} + ~O(\alpha^2(Q^2)) 
\nonumber \\
&\approx & 
\lambda^{eff,as}_q(z,Q^2) + \frac{3 \alpha(Q^2)}{\ln(1/x)},
\nonumber
\end{eqnarray}
where
symbol $\approx $ marks approximations obtained by expansions of modified
Bessel functions $I_n(\sigma)$. The slope 
$\lambda^{eff,as}_{F2}(x,Q^2)$ lies between quark and gluon ones but
closely to quark slope $\lambda^{eff,as}_{q}(x,Q^2)$ (see also Fig. 2).

\item
Both slopes $\lambda^{eff}_a$ decrease with decreasing $x$. 
A $x$ dependence of the slope should not appear
for a PD with a Regge type
asymptotic ($x^{-\lambda}$) and precise measurement of the slope 
$\lambda^{eff}_a$ may lead to the possibility to verify the type of small
$x$ asymptotics of parton distributions.
\end{itemize}

%

%\section{Results of the fits} \indent

{\bf 2.} With the help of the results obtained in the previous section we have
analyzed $F_2$ HERA data at small $x$ from the H1 collaboration
(first article in \cite{H1}).
%and ZEUS \cite{ZEUS} collaborations separately.
%Initially our solution of the DGLAP equations depends on five
%parameters, i.e. $Q_0^2$, $x_0$, $A_q$, $A_g$ and $\Lambda_{\ms}(n_f=4)$.
In order to keep the analysis as simple as possible
we have fixed $\Lambda_{\ms}(n_f=4) = 250$ MeV which
is a reasonable value extracted from the traditional (higher $x$)
experiments.
%and that has also been used by others \cite{Munich1}.
The initial scale of the PD was also fixed
into the fits to $Q^2_0$ = 1 $GeV^2$, although later it was released
to study the sensitivity of the fit to the variation of this parameter.
The analyzed data region was restricted to $x<0.01$ to remain within the
kinematical range where our results are
accurate. Finally, the number of active flavors was fixed to $f$=4. 

Fig. 1 shows $F_2$ calculated from the fit
with Q$^2$ $>$ 1 GeV$^2$
%given in table 1
in comparison with H1 data.
Only the lower $Q^2$ bins are shown.
One can observe that the NLO result (dot-dashed line)
lies closer to the data
than the LO curve (dashed line).
The lack of agreement between data and lines observed
at the lowest $x$ and $Q^2$ bins suggests
that the flat behavior should occur at $Q^2$ lower
than 1 GeV$^2$.
In order to study this point we have done the
analysis considering $Q_0^2$ as a free parameter.
Comparing the results of the fits (see \cite{Q2evo})
%in table 3 with those in table 2
one can notice
%a significant reduction in the value of
%$A_g$, $Q_0^2$ and the $\chi^2$. In Fig. 1
the better agreement with the experiment of the
NLO curve at fitted $Q^2_0=0.55 GeV^2$ (solid curve)
is apparent at the lowest kinematical bins.

Finally with the help of Eqs. (\ref{11.1}) we have estimated
the $F_2$ effective slope using the value of the parameters extracted
from NLO fits to data. For H1 data we found
$0.05 < \lambda^{eff}_{F2} < 0.30-0.37$.
%and for ZEUS   
%$0.07-0.09 < \lambda^{eff}_{F2} < 0.31-0.34$.
The lower (upper) limits
corresponds \footnote{For small $Q^2$ we used the exact values of the
slopes presented in \cite{Q2evo}.}
to $Q^2=1.5$ GeV$^2$ ($Q^2=400$ GeV$^2$). The dispersion
in some of the limits is due to the $x$ dependence.
Fig. 2 shows that the three types of asymptotical slopes have
similar values,
which are in very good agreement with H1 data (presented also in Fig. 2).
The NLO values of
$\lambda^{eff,as}_{F2}$ lie between the quark and the gluon ones but
closer to the quark slope $\lambda^{eff,as}_{q}$.
These results are in excellent agreement with those obtained by others 
(see %references \cite{H1,MRS,GRV,Navelet} and also
the review
\cite{CoDeRo} and references therein).\\

%\section{Conclusions} \indent
{\bf 3.} {\it Resume.}
We have shown that the results developed recently in \cite{Q2evo}
%As we have shown, these results
 have quite simple form and reproduce many
properties of parton distributions at small $x$,
that have been known from global fits.

We found very good agreement between our approach based on QCD at
NLO approximation and HERA data, as it has been observed earlier with
other approaches (see the review \cite{CoDeRo}). Thus, the nonperturbative
contributions as shadowing effects,
%\cite{Levin},
higher twist effects
%\cite{Bartels}
and others seems to be quite small or seems to be canceled
between them and/or with $ln(1/x)$ terms containing by higher orders of
perturbative theory. To clear up the correct contributions of nonperturbative
dynamics and higher orders containing strong $ln(1/x)$ terms, it is
necessary
more precise data and further efforts in developing of theoretical
approaches.\\
%

%
%\vspace{1cm}
%\hspace{1cm} \Large{} {\bf Acknowledgements}    \vspace{0.5cm}

%\normalsize{}

%
One of the authors (G.P.) was supported in part by Xunta de Galicia
(XUGA-20604A96) and CICYT (AEN96-1673).

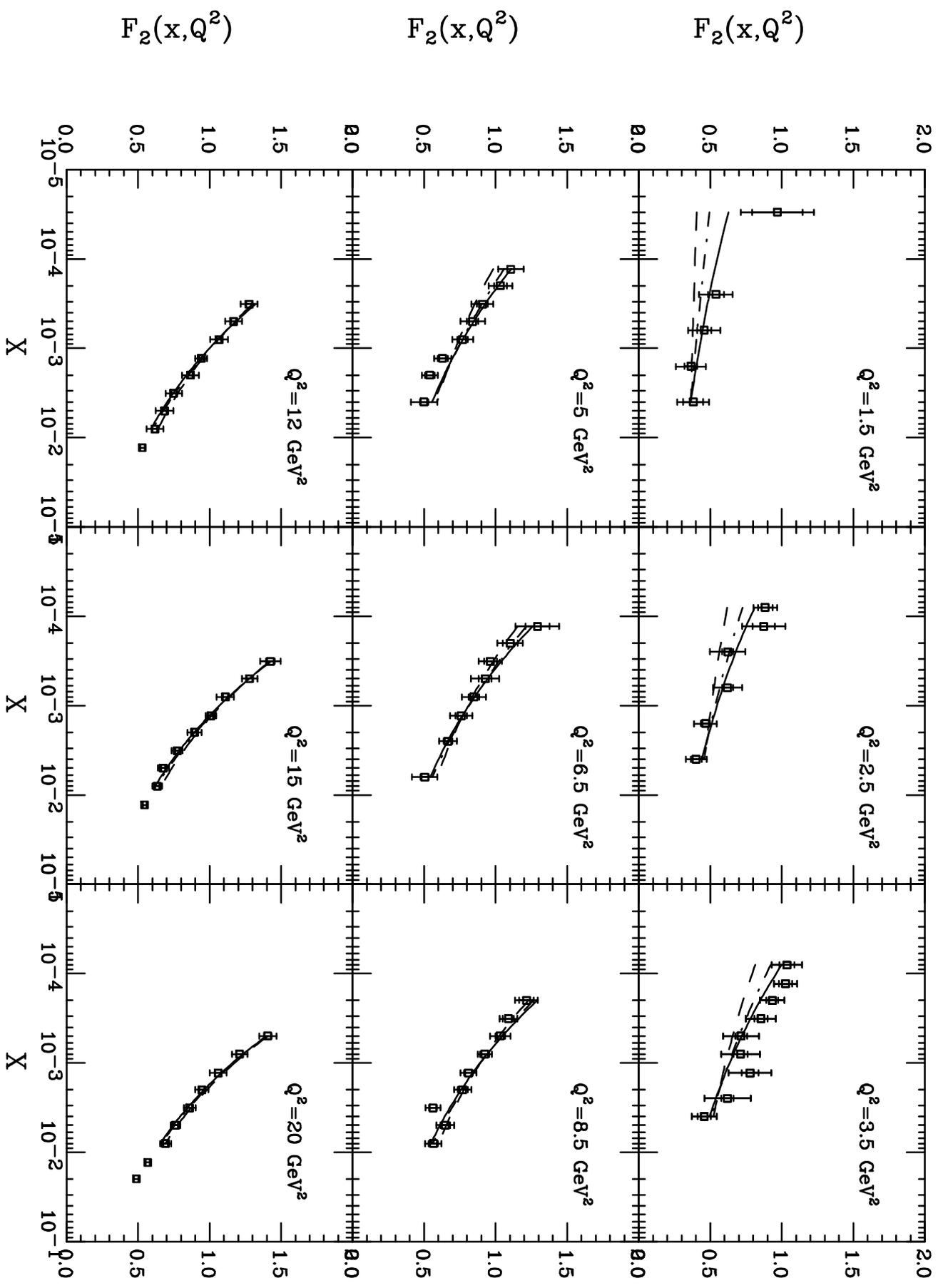
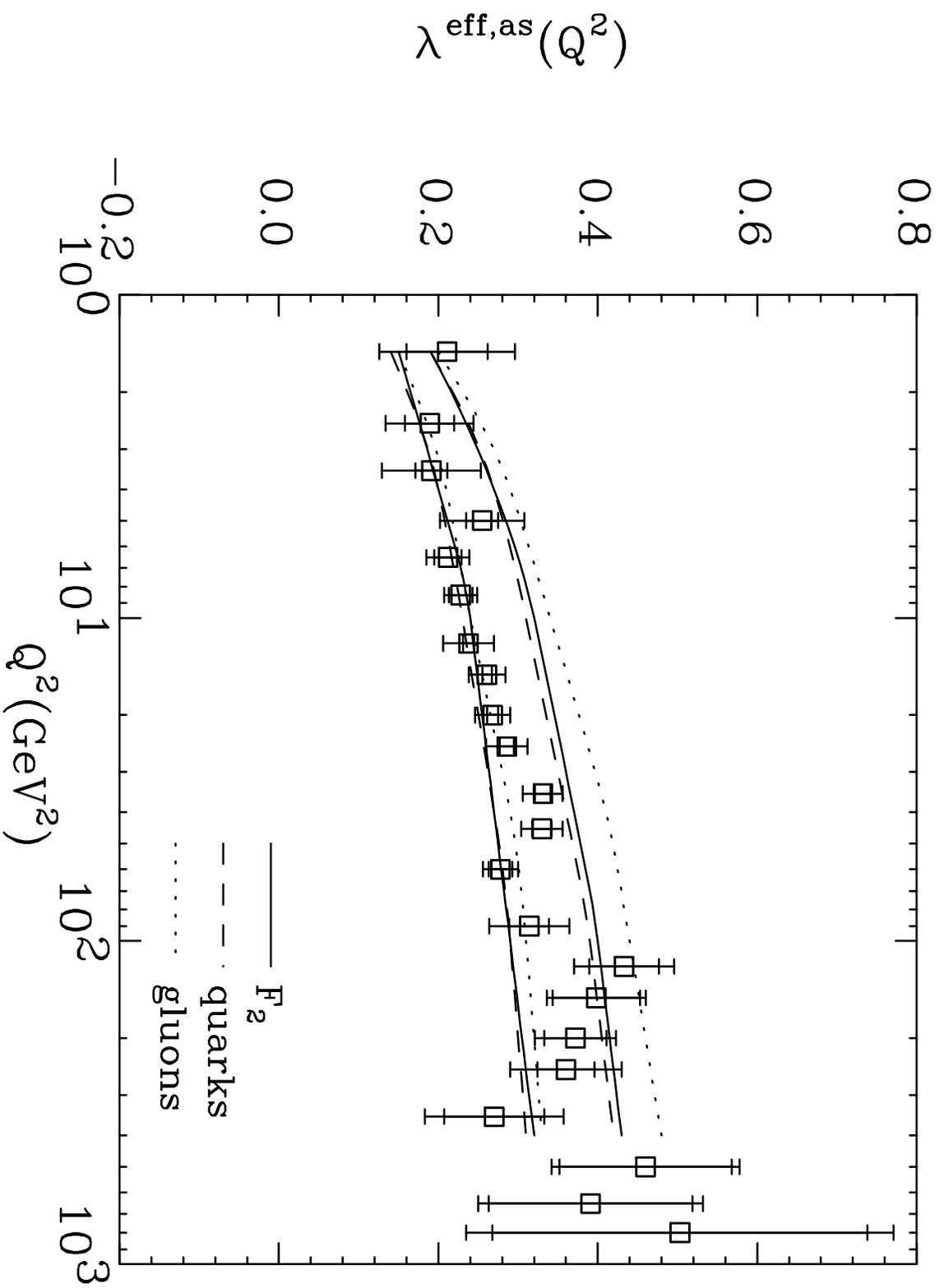
\end{document}